\documentclass[12pt,aps,prd,showpacs,groupedaddress,notitlepage,nofootinbib,preprintnumbers]{revtex4-1}

\usepackage{amsmath,amssymb,amsfonts}
\usepackage{bm}
\usepackage{color,graphicx}

\newcommand{\llangle}{\langle\!\langle}
\newcommand{\rrangle}{\rangle\!\rangle}
\newcommand{\0}{\bm{0}}
\newcommand{\1}{\bm{1}}
\newcommand{\I}{\bm{i}}
\newcommand{\J}{\bm{j}}
\newcommand{\K}{\bm{k}}
\newcommand{\EP}{\bm{e}_+}
\newcommand{\EM}{\bm{e}_-}
\newcommand{\EPM}{\bm{e}_{\pm}}
\newcommand{\EMP}{\bm{e}_{\mp}}

\begin{document}


\title{Energy-parity from a bicomplex algebra}
\author{Max Edward Laycock}
\author{Peter Millington}
\email{Corresponding author: p.millington@nottingham.ac.uk}
\affiliation{School of Physics and Astronomy, University of Nottingham, Nottingham NG7 2RD, United Kingdom}
\date{\today}


\begin{abstract}
\vspace{2em}
By replacing the field of complex numbers with the commutative ring of bicomplex numbers, we attempt to construct interacting scalar quantum field theories that feature both positive- and negative-energy states. This work places the tentative ideas proposed in [R.~Dickinson, J.~Forshaw and P.~Millington, J.\ Phys.\ Conf.\ Ser.\  {\bf 631} (2015) 012059] on more solid and general mathematical foundations and incorporates the ``energy-parity" symmetry introduced in [A.~D.~Linde, Phys.\ Lett.\ B {\bf 200} (1988) 272; D.~E.~Kaplan and R.~Sundrum, JHEP {\bf 0607} (2006) 042]. The interplay of the positive- and negative-energy states allows for cancellations of the vacuum energy. Both the positive- and negative-energy states have positive norms, and their direct mixing is prevented by virtue of the zero divisors of the bicomplex numbers, thereby eliminating the possibility of negative-energy cascades.  We suggest that the same interplay of positive- and negative-energy states may allow Haag's theorem to be circumvented, removing the associated criticism of the Fock representation. We consider scalar theories with cubic and quartic interactions and describe how this construction may yield transition probabilities consistent with the standard scattering theory. Whilst these results are intriguing, we draw attention to potentially serious limitations in relation to perturbative unitarity.
\end{abstract}


\maketitle

\newpage


\section{Introduction}

In spite of the successes of the Standard Model of particle physics and the concordance cosmology, fundamental physics finds itself at an uncomfortable juncture. This discomfort stems from the naturalness problem and our theoretical discontent with both the smallness of the cosmological constant and the disparity between the electroweak and Planck scales. These obstacles have at their origins the behaviour of the vacuum in quantum field theory, and the latter is the focus of this article.

In the Fock representation of the interaction picture, the Minkowski vacuum has a curious property: since the Fock space is spanned by a complete basis of states of positive energy, any vacuum fluctuation is biased to have a net positive energy (again in Minkowski spacetime). This can be seen concretely by considering the evolution of the interaction-picture vacuum state $|0\rangle\equiv |\Omega(t=0)\rangle$ under the influence of an interaction Hamiltonian $\hat{H}^{\rm int}(t)$:
\begin{equation}
  |\Omega(t)\rangle\ =\ \hat{U}(t,0)|0\rangle\;.
\end{equation}
The (improper) unitary evolution operator
\begin{equation}
  \hat{U}(t,0)\ =\ T\,\exp\Bigg[-\,i\int_0^t\!{\rm d}t'\;\hat{H}^{\rm int}(t')\Bigg]
\end{equation}
where $T$ indicates time-ordering, can be projected into the Fock basis by completeness, such that all vacuum loops are built out of modes of positive energy. We recall that off-shell effects emerge in the interaction picture by virtue of the time-ordering operation and the resulting convolution of expectation values of field operators with unit step functions. It is, perhaps, unsurprising then that vacuum fluctuations are problematic in quantum field theory: we cannot expect the expectation value of the energy of the Minkowski vacuum to be zero, if its quantum fluctuations are biased so as always to increase the vacuum energy towards the ultraviolet. It is not unreasonable to speculate that this is, in fact, the root of Haag's critique of the Fock representation~\cite{Haag:1955ev}.

There is a long history of attempts to rectify this curious behaviour by introducing states of \emph{negative} energy, whose contributions allow for downward fluctuations in the vacuum energy. A notable example is the ``energy-parity'' symmetry of Kaplan and Sundrum~\cite{Kaplan:2005rr} (see also~Ref.~\cite{Elze:2005kt}) and Linde's Universe multiplication~\cite{Linde:1988ws} (for an overview, see Ref.~\cite{Padilla:2015aaa}), and similar ideas for removing the ultraviolet sensitivity of quantum field theory based on indefinite inner-product spaces go back as far as early works by Dirac~\cite{Dirac} and Pauli~\cite{Pauli}. However, `ghost' constructions often cannot prevent mixing between the positive- and negative-energy degrees of freedom when coupled to gravity, and the resulting negative-energy cascades destabilize the vacuum. A similar cancellation of the vacuum energy occurs in the suggestion of 't Hooft and Nobbenhuis~\cite{tHooft:2006uhw}, wherein the negative-energy operators are obtained from the positive-energy ones by an analytic continuation $x^{\mu}\to ix^{\mu}$. However, for massive fields, this procedure leads to negative-energy states with tachyonic mass terms.

The central idea of this work is to abandon the algebraic property of division in constructing quantum field theory, by replacing the field of complex numbers by the commutative ring of bicomplex numbers. In doing so, we will be able to partition the algebra of our quantum field theory into two ideal subalgebras, associating one with positive-energy states and the other with negative-energy states. The intersection of these two subalgebras is a singleton, containing only the zero element, and it will turn out to play a privileged role in constructing the vacuum state of the quantum field theory. The interplay of the positive- and negative-energy states leads to a cancellation of the zero-point energy.  However, in contrast to `ghost' constructions, the complementarity of the positive- and negative-energy subalgebras ensures that there can be no direct mixing of the associated modes, such that negative-energy cascades are avoided. Moreover, the same interplay may allow the Fock representation to avoid the criticisms of Haag's theorem~\cite{Haag:1955ev}. Probabilities are associated with the Euclidean inner product of the bicomplex numbers, and both the positive- and negative-energy Fock states have positive norms with respect to this inner product.

The construction presented here takes its inspiration from the tentative ideas presented in Ref.~\cite{Dickinson:2015ixa}. Therein, the ring of bicomplex numbers was used to make an energy-parity symmetry manifest, and the sensitivity of the resulting theory to vacuum fluctuations was studied. The results presented here bear strong resemblance to those of Ref.~\cite{Dickinson:2015ixa}, with the important exception that we do not have to exclude direct couplings between the positive- and negative-energy states `by hand'~\cite{Dickinson:2015ixa}; this separation is instead a direct consequence of the algebraic structure. The potential implications of hypercomplex algebras for the sensitivity of quantum field theory to vacuum fluctuations has also been acknowledged recently in Refs.~\cite{Cartas-Fuentevilla:2016ccg,Cartas-Fuentevilla:2017ged}.

Whilst we find the present construction to have a number of intriguing properties, it is not without potentially serious limitations, many of which were originally noted in Ref.~\cite{Dickinson:2015ixa} and which we do not address here. Not least of all, it remains to be seen whether the theories permitted by this construction are consistent with perturbative unitarity.

The remainder of this article is organized as follows. In Sec.~\ref{sec:algebra}, we provide an introduction to the algebra of the bicomplex numbers, which forms the basis of our subsequent constructions. We then proceed in Sec.~\ref{sec:scalar} to describe a free scalar quantum field theory that incorporates both positive- and negative-energy states, illustrating how the zero-point energy is eliminated. In Sec.~\ref{sec:Haag}, we suggest that this construction may avoid the criticisms of Haag's theorem.  We consider interacting theories in Sec.~\ref{sec:inttheory} and describe how they may yield transition probabilities consistent with standard results. Speculative remarks in the context of gravity are given in Sec.~\ref{sec:gravity}, and our conclusions are presented in Sec.~\ref{sec:conclusions}.


\section{Algebraic structure}
\label{sec:algebra}

The field of complex numbers $\mathbb{C}(\I)$ ($\I^2=-\1$) forms a commutative division algebra. It is an Abelian group under both addition and multiplication, and each element has a unique additive and multiplicative inverse (with the exception of the zero element). By introducing a second, distinct imaginary unit $\J$ ($\J^2=-\1$), we can construct the ring of bicomplex numbers $\mathbb{BC}$, which forms a four-dimensional associative and commutative algebra over the real numbers, spanned by the basis $\{\bm{1},\I,\J,\K\equiv \I\J\}$. Its algebra is an Abelian group under addition but a monoid under multiplication, and there does not exist a multiplicative inverse for each non-zero element. Therefore, unlike the more familiar complex numbers, the bicomplex numbers do not form a division algebra. In fact, the algebra of the bicomplex numbers is isomorphic to the Clifford algebra $C\ell_1(\mathbb{C})$, the unique commutative complex Clifford algebra that is not also a division algebra.

The bicomplex algebra is equipped with three distinct automorphisms: complex conjugation with respect to $\I$, which we denote by $\ast$; complex conjugation with respect to $\J$, which we denote by $\star$; and bicomplex conjugation $\times\equiv \ast\circ\star$. Under their actions, we have
\begin{subequations}
\begin{gather}
  \I^{\ast}\ =\ -\:\I\;,\qquad \J^{\ast}\ =\ +\:\J\;,\qquad \K^{\ast}\ =\ -\:\K\;,\\
  \I^{\star}\ =\ +\:\I\;,\qquad \J^{\star}\ =\ -\:\J\;,\qquad \K^{\star}\ =\ -\:\K\;,\\
  \I^{\times}\ =\ -\:\I\;,\qquad \J^{\times}\ =\ -\:\J\;,\qquad \K^{\times}\ =\ +\:\K\;.
\end{gather}
\end{subequations}
We can therefore define three distinct quadratic forms for a bicomplex number $X\in \mathbb{BC}$, each associated with one of the subalgebras $\mathbb{C}(\I)$, $\mathbb{C}(\J)$ or the duplex numbers $\mathbb{D}(\K)$ (spanned by $\bm{1}$ and the hyperbolic unit $\K$, for which $\K^2=\1$):
\begin{align}
  \label{eq:invCi}
  {|X|}^{2}_{\mathbb{C}(\I)}\  &\equiv\  X{X}^{*} \ \in \ \mathbb{C}(\J)\;,\\
  \label{eq:invCj}
  {|X|}^{2}_{\mathbb{C}(\J)} \ &\equiv\ X{X}^{\star} \ \in \ \mathbb{C}(\I)\;,\\
  \label{eq:invD}
  {|X|}^{2}_{\mathbb{D}(\K)} \ &\equiv\  X{X}^{\times} \ \in \ \mathbb{D}(\K)\;.
\end{align}

The bicomplex numbers include two complementary (orthogonal) and idempotent zero divisors:
\begin{equation}
  \EPM\ \equiv\ \frac{1}{2}\,\big(\1\pm\K\big) \in\ \mathbb{D}(\K)\;,
\end{equation}
satisfying
\begin{equation}
  \label{eq:defs}
  \EPM^2\ =\ \EPM\;,\qquad \qquad \EP\:+\:\EM\ =\ \1\;,\qquad \qquad \EP\:-\:\EM\ =\ \K\;,
\end{equation}
with
\begin{equation}
  \EP\cdot \EM\ =\ \0\;,\qquad \qquad \K\cdot \EPM\ =\ \pm\,\EPM\;.
\end{equation}
It follows that the two subalgebras $\mathbb{BC}_{\pm}\equiv \EPM\cdot\mathbb{BC}$ are principal ideals of the algebra, whose intersection is the zero element:
\begin{equation}
  \mathbb{BC}_+\ \cap\ \mathbb{BC}_-\ =\ \{0\}\;.
\end{equation}
We will see later that this singlet intersection will play a special role in defining the vacuum state of our quantum field theory.

Making use of the zero divisors, we can introduce the so-called idempotent decomposition,  which allows us to write a bicomplex number $X$ in any of the following equivalent forms:
\begin{align}
  X\ &=\ \EP \cdot X\:+\:\EM \cdot X\nonumber\\ &=\ X_+\:+\:X_-\nonumber\\ &
  =\ \EP \cdot x^{(\I)}_+\:+\:\EM \cdot x^{(\I)}_-\nonumber\\ &=\ \EP \cdot x^{(\J)}_+\:+\:\EM \cdot x^{(\J)}_-\;,
\end{align}
where $X_{\pm}\equiv \EPM\cdot X\in\mathbb{BC}_{\pm}$, $x_{\pm}^{(\I)}\in\mathbb{C}(\I)$ and $x^{(\J)}_{\pm}\in\mathbb{C}(\J)$. We can therefore associate any bicomplex number with an ordered pair \smash{$(x^{(\I)}_+,x^{(\I)}_-)$} of $\mathbb{C}(\I)$ complex numbers or \smash{$(x^{(\J)}_+,x^{(\J)}_-)$} of $\mathbb{C}(\J)$ complex numbers, and this reflects the isomorphism $\mathbb{BC}\simeq \mathbb{C}_2(i)\simeq \mathbb{C}_2(j)$. The basis \smash{$\{\EP, \EM\}$} is orthogonal (but not orthonormal) with respect to the inner product of the linear vector space $\mathbb{C}_2(i)$, and we can define the Euclidean inner products
\begin{equation}
  \langle{ \EP,\EM}\rangle_{\mathbb{C}_2(i)} \ =\  0\;,\qquad \qquad \langle{ \EPM,\EPM}\rangle_{\mathbb{C}_2(i)}\
  =\ \frac{1}{2}\;,
\end{equation}
where the zero divisors have the $\mathbb{C}_2(i)$ linear vector space representations
\begin{equation}
  \EPM\ \simeq\ \frac{1}{2}\begin{pmatrix}1 & \mp\,i\\ \pm\,i & 1\end{pmatrix}\;.
\end{equation}
The Euclidean inner product will allow us to obtain real-valued probabilities, and the fact that the zero divisors do not have unit norm with respect to this Euclidean inner product will turn out to be important for ensuring that we obtain the correct normalization of transition probabilities.

For the discussions that follow, it is helpful to introduce the Dirac notation
\begin{equation}
  |\EPM\rangle_{\mathbb{C}_2(i)}\ \simeq\ \EPM\cdot\begin{pmatrix}1 \\ 0\end{pmatrix}\ =\ \frac{1}{2}\begin{pmatrix} 1 \\ \pm\,i\end{pmatrix}\;,
\end{equation}
where the vectors satisfy
\begin{equation}
  \EPM|\EPM\rangle_{\mathbb{C}_2(i)}\ =\ |\EPM\rangle_{\mathbb{C}_2(i)}\;,\qquad \qquad \K|\EPM\rangle_{\mathbb{C}_2(i)}\ =\ \pm\,|\EPM\rangle_{\mathbb{C}_2(i)}
\end{equation}
We also make note of the linear operators
\begin{align}
  |\EPM\rangle_{\mathbb{C}_2(i)}\langle\EPM|_{\mathbb{C}_2(i)}\ &=\ \frac{1}{4}\begin{pmatrix} 1 & \mp\, i \\ \pm\,i & 1\end{pmatrix}\;,\\
  |\EPM\rangle_{\mathbb{C}_2(i)}\langle\EMP|_{\mathbb{C}_2(i)}\ &=\ \frac{1}{4}\begin{pmatrix} 1 & \mp\, i \\ \mp\,i & -\,1\end{pmatrix}\;.
\end{align}
The $\mathbb{C}_2(i)$ linear vector space representations of $\1$ and $\K$ are
\begin{equation}
  \1\ =\ \begin{pmatrix} 1 & 0 \\ 0 & 1\end{pmatrix}\;,\qquad
  \K\ =\ \begin{pmatrix} 0 & -\,i \\ +\,i & 0\end{pmatrix}\;,
\end{equation}
such that
\begin{equation}
  |\1\rangle_{\mathbb{C}_2(i)}\ =\ \begin{pmatrix} 1 \\ 0 \end{pmatrix}\;,\qquad
  |\K\rangle_{\mathbb{C}_2(i)}\ =\ \begin{pmatrix} 0 \\ i\end{pmatrix}\;,
\end{equation}
and, for completeness, those of $\I$ and $\J$ are
\begin{equation}
  \I\ =\ \begin{pmatrix} i & 0 \\ 0 & i\end{pmatrix}\;,\qquad
  \J\ =\ \begin{pmatrix} 0 &+\, 1 \\ -\,1 & 0\end{pmatrix}\;.
\end{equation}
For later reference, we note that the square-root of $\K$ can be written as
\begin{equation}
  \label{eq:rootk}
  \K^{1/2}\ =\ \frac{\K^n}{\sqrt{2}}\,e^{i(\pi/4+m\pi)}\begin{pmatrix} 1 & -\,1 \\ 1 & 1\end{pmatrix}\;,\qquad n,\:m\ =\ 0,\: 1\;.
\end{equation}

Employing the isomorphism $\mathbb{BC}\simeq\mathbb{C}_2(i)$ and given two bicomplex numbers $X$ and $Y$, their Euclidean inner product is
\begin{equation}
  \langle{ X,Y}\rangle_{\mathbb{C}_2(i)}\ =\ \frac{1}{2}\,\Big(\langle x_+,y_+\rangle_{\mathbb{C}(i)}\:
  +\:\langle x_-,y_-\rangle_{\mathbb{C}(i)}\Big)\;,
\end{equation}
where
\begin{equation}
  \langle x,y\rangle_{\mathbb{C}(i)}\ =\ x^*y\;.
\end{equation}
The associated Euclidean norm is real-valued and positive semi-definite:
\begin{equation}
  |X|_{\mathbb{C}_2(i)}^2\ =\ \frac{1}{2}\Big(|x_+|_{\mathbb{C}(i)}^2\:+\:|x_-|^2_{\mathbb{C}(i)}\Big)\ \geq\ 0\;.
\end{equation}

The above structures can readily be extended to a bicomplex module $M$. Defining the submodules $M_+\equiv \EP\cdot M$ and $M_-\equiv \EM\cdot M$, we have $M_+\cap M_-=\{0\}$ and any bicomplex module $M$ can be represented by the orthogonal direct sum $M=M_+[\oplus] M_-$. For example, given two $n$-dimensional bicomplex vectors $U=\EP\cdot u_++\EM\cdot u_-\in M$ and $V=\EP\cdot v_++\EM\cdot v_-\in M$, we have
\begin{align}
  \llangle U,V\rrangle\ \equiv\ \langle U,V\rangle_{\mathbb{C}^n_2(i)}\ &
  =\ \langle \EP\cdot u_+,\EP\cdot v_+\rangle_{\mathbb{C}^n_2(i)}\:+\:\langle \EM \cdot u_-,\EM\cdot v_-\rangle_{\mathbb{C}^n_2(i)}
  \nonumber\\ &= \ \frac{1}{2}\Big(\langle u_+,v_+\rangle_{\mathbb{C}^n(i)}\:+\:\langle u_-,v_-\rangle_{\mathbb{C}^n(i)}\Big)\;,
\end{align}
where
\begin{equation}
  \langle u,v\rangle_{\mathbb{C}^n(i)}\ =\ u^{\dag}v
\end{equation}
and $\dagger$ denotes the usual Hermitian conjugate (involving complex conjugation with respect to $i$ and matrix transposition). Moreover, a linear map $L:M\longrightarrow M'$, where $M$ and $M'$ are $\mathbb{BC}$-modules, can be written in terms of mappings $L_{\pm}:M_{\pm}\longrightarrow M_{\pm}'$ that act on each of the submodules, such that any linear operator on a bicomplex module can also be written in the idempotent decomposition $L=\EP\cdot L+\EM\cdot L=L_++L_-$.


\section{Free scalar theory}
\label{sec:scalar}

We begin with the bicomplex Hilbert module $M=M_+[\oplus]M_-$, and associate the submodule $M_+$ with the usual Hilbert space $\mathcal{H}$ (over $\mathbb{C}(i)$) of the massive scalar quantum field theory with Lagrangian
\begin{equation}
  \mathcal{L}\ =\ \frac{1}{2}\,\partial_{\mu}\phi(x)\,\partial^{\mu}\phi(x)\:-\:\frac{1}{2}\,m^2\,\phi^2(x)\;,
\end{equation}
i.e.~we define $M_+\equiv\EP\cdot \mathcal{H}$. We work with the signature convention $(-,+,+,+)$ throughout this article. In the interaction picture, the canonical algebra
\begin{subequations}
\begin{align}
  [\hat{\phi}(t,\mathbf{x}),\hat{\phi}(t,\mathbf{y})]\ &=\ 0\;,\\
  [\hat{\pi}(t,\mathbf{x}),\hat{\phi}(t,\mathbf{y})]\ &=\ i\,\delta^3(\mathbf{x}-\mathbf{y})\;,
\end{align}
\end{subequations}
where $\hat{\pi}(t,\mathbf{x})\equiv \,\dot{\!\hat{\phi}}(t,\mathbf{x})$ is the conjugate-momentum operator, can be formulated in terms of the single-particle annihilation operator $\hat{a}(t,\mathbf{p})\in\mathcal{H}$ and its Hermitian conjugate $\hat{a}^{\dag}(t,\mathbf{p})\in\mathcal{H}$. These operators satisfy the canonical commutation relation
\begin{equation}
  \label{eq:CCR}
  [\hat{a}(t,\mathbf{p}),\hat{a}^{\dag}(t,\mathbf{p}')]\ =\ (2\pi)^32E(\mathbf{p})\delta^3(\mathbf{p}-\mathbf{p}')\;,
\end{equation}
where $E(\mathbf{p})=\sqrt{\mathbf{p}^2+m^2}$ is the on-shell energy, and all other commutators vanish. The creation operator $\hat{a}^{\dag}(t,\mathbf{p})$ acts on the Fock-space vacuum state $|0\rangle$ to produce a positive-energy, single-particle state of momentum $\mathbf{p}$:
\begin{equation}
  \hat{a}^{\dag}(t,\mathbf{p})|0\rangle\ =\ |t,\mathbf{p}\rangle\ =\ e^{iE(\mathbf{p})t}|t=0,\mathbf{p}\rangle\;,
\end{equation}
which is annihilated by the operator $\hat{a}(t,\mathbf{p})$, i.e.
\begin{equation}
  \hat{a}(t,\mathbf{p})|t',\mathbf{p}'\rangle\ =\ e^{iE(\mathbf{p})(t'-t)}(2\pi)^32E(\mathbf{p})\delta^3(\mathbf{p}-\mathbf{p}')|0\rangle\;.
\end{equation}

The annihilation operator $\hat{a}(t,\mathbf{p})$ can be embedded in the submodule $M_+$ as
\begin{equation}
  \label{eq:a1}
  \hat{a}_+(t,\mathbf{p})\ \equiv\ \EP\otimes\hat{a}(t,\mathbf{p})\;,
\end{equation}
and the creation operator is the bicomplex adjoint (denoted by $\#$ and corresponding to the bicomplex conjugate of the operator transpose)
\begin{equation}
  \label{eq:a2}
  \hat{a}^{\#}_+(t,\mathbf{p})\ \equiv\ \EP\otimes\hat{a}^{\dag}(t,\mathbf{p})\;.
\end{equation}
In order to obtain real-valued probabilities, expectation values will be taken with respect to the Euclidean inner product of the bicomplex numbers, and the Kronecker products in Eqs.~\eqref{eq:a1} and \eqref{eq:a2} indicate that the zero divisors $\EPM$ are to be understood in the $\mathbb{C}_2(i)$ linear vector space representation discussed in Sec.~\ref{sec:algebra}. For convenience, we will hereafter employ a simplified notation in which the Kronecker product and unit operators in $\mathcal{H}$ are omitted, e.g.~$\EPM\equiv \EPM\:\otimes\:\hat{\mathbb{I}}$.  The canonical commutation relation in Eq.~\eqref{eq:CCR} is translated to
\begin{equation}
  [\hat{a}_+(t,\mathbf{p}),\hat{a}_+^{\#}(t,\mathbf{p}')]\ =\ (2\pi)^32E(\mathbf{p})\delta^3(\mathbf{p}-\mathbf{p}')\,\EP\;,
\end{equation}
with all other commutators of positive-energy operators vanishing. 

We may now introduce \emph{negative}-energy states by defining the submodule $M_-$ in terms of the Hilbert space spanned by the time-reversed states, i.e.~$M_-\equiv M_+^*$. We embed the relevant annihilation and creation operators as
\begin{subequations}
\begin{align}
  \hat{a}_-(t,\mathbf{p})\ &\equiv\ \EM\,\hat{a}^*(t,\mathbf{p})\;,\\
  \hat{a}_-^{\#}(t,\mathbf{p})\ &=\ \EM\,\hat{a}^{\mathsf{T}}(t,\mathbf{p})\;,
\end{align}
\end{subequations}
where $\mathsf{T}$ denotes the operator transpose. These operators satisfy the commutation relation
\begin{equation}
  [\hat{a}_-(t,\mathbf{p}),\hat{a}_-^{\#}(t,\mathbf{p}')]\ =\ (2\pi)^32E(\mathbf{p})\delta^3(\mathbf{p}-\mathbf{p}')\,\EM\;,
\end{equation}
with all other commutators of negative-energy operators vanishing. Moreover, by virtue of the orthogonality of the zero divisors $\EP$ and $\EM$, all commutators of $+$ and $-$ operators also vanish, i.e.
\begin{subequations}
\begin{align}
  [\hat{a}_{\pm}(t,\mathbf{p}),\hat{a}_{\mp}(t,\mathbf{p}')]\ &=\ \0\;,\\
  [\hat{a}_{\pm}(t,\mathbf{p}),\hat{a}^{\#}_{\mp}(t,\mathbf{p}')]\ &=\ \0\;.
\end{align}
\end{subequations}

In terms of the vacuum state of the usual Fock space $|0\rangle\in\mathcal{H}$, which is self-conjugate under time-reversal, i.e.~$|0\rangle\in\mathcal{H}^*$, the vacuum states for the positive and negative operators are $|0_+\rrangle\equiv |\EP\rangle\otimes|0\rangle$ and $|0_-\rrangle\equiv |\EM\rangle\otimes|0\rangle$. We again omit the subscript $\mathbb{C}_2(i)$ in the linear vector space representations of $|\EPM\rangle$ for notational simplicity. It follows that the unique element which forms the intersection of the submodules $M_+$ and $M_-$ is just the vacuum state of the original Hilbert space, i.e.
\begin{equation}
  |0\rrangle\ \equiv\ |\1\rangle\otimes|0\rangle\ \ =\ (|\EP\rangle+|\EM\rangle)\otimes|0\rangle\ 
  =\ |0_+\rrangle\:+\:|0_-\rrangle\ \in \ M_+\:\cap\: M_-\;.
\end{equation}
We can now introduce positive- and negative-energy, single-particle Fock states $|t,\mathbf{p}_{\pm}\rrangle$ through the following actions of the creation and annihilation operators:
\begin{align}
  \hat{a}_{\pm}^{\#}(t,\mathbf{p})|0\rrangle\ &=\ e^{\pm \I E(\mathbf{p})t}|t=0,\mathbf{p}_{\pm}\rrangle\;,\\
  \hat{a}_{\pm}(t,\mathbf{p})|t',\mathbf{p}_{\pm}'\rrangle\ &
  =\ e^{\pm \I E(\mathbf{p})(t'-t)}(2\pi)^32E(\mathbf{p})\delta^3(\mathbf{p}-\mathbf{p}')|0_{\pm}\rrangle\;.
\end{align}
It immediately follows that these states are orthogonal with respect to the Euclidean inner product:
\begin{align}
  \label{eq:Fockstates1}
  \llangle \mathbf{p}_{\pm}'|\mathbf{p}_{\pm}\rrangle\ &=\ (2\pi)^3E(\mathbf{p})\delta^3(\mathbf{p}-\mathbf{p}')\;,\\
  \label{eq:Fockstates2}
  \llangle \mathbf{p}_{\pm}'|\mathbf{p}_{\mp}\rrangle\ &=\ 0\;.
\end{align}
Notice that both the positive- and negative-energy states have positive-definite norms. An $n$-particle state takes the form
\begin{equation}
  |t,\mathbf{p}_{1\pm};\mathbf{p}_{2\pm};\dots;\mathbf{p}_{n\pm}\rrangle\
  =\ |\EPM\rangle\otimes|t,\mathbf{p}_{1};\mathbf{p}_{2};\dots;\mathbf{p}_{n}\rangle^{(*)}\ 
  =\ \hat{a}_{\pm}^{\#}(t,\mathbf{p}_1)\hat{a}_{\pm}^{\#}(t,\mathbf{p}_2)\dots \hat{a}_{\pm}^{\#}(t,\mathbf{p}_n)|0\rrangle\;,
\end{equation}
and we draw attention to the normalization of the states, which differ from the standard ones by an overall factor of 2, due to the normalization of the idempotents $\EPM$. With this difference in normalization, the completeness of the Fock space takes the form
\begin{equation}
  \hat{\mathbb{I}}\ =\ \hat{\mathbb{I}}_+\:+\:\hat{\mathbb{I}}_-\;,
\end{equation}
where
\begin{align}
  \mathbb{I}_{\pm}\ &\equiv\ 2\bigg[|0_{\pm}\rrangle \llangle 0_{\pm}|\:
  +\:\int\!{\rm d}\Pi_{\mathbf{p}}\;|0,\mathbf{p}_{\pm}\rrangle\llangle 0,\mathbf{p}_{\pm}|\nonumber\\&\qquad
  +\:\frac{1}{2!}\int\!{\rm d}\Pi_{\mathbf{p}}\,{\rm d}\Pi_{\mathbf{p}'}\;|0,\mathbf{p}_{\pm};0,\mathbf{p}_{\pm}'\rrangle
  \llangle 0,\mathbf{p}_{\pm};0,\mathbf{p}_{\pm}'|\:+\:\dots\bigg]\;,
\end{align}
and we use the shorthand notation
\begin{equation}
  \int\!{\rm d}\Pi_{\mathbf{p}}\ \equiv\ \int\!\frac{{\rm d}^3\mathbf{p}}{(2\pi)^3}\,\frac{1}{2E(\mathbf{p})} 
\end{equation}
for the Lorentz-invariant phase-space measure.

The Heisenberg equation for the annihilation operators takes the form
\begin{equation}
  \frac{\rm d}{{\rm d}t}\,\hat{a}_{\pm}(t,\mathbf{p})\ =\ \I\big[\hat{\bm{H}}^0,\hat{a}_{\pm}(t,\mathbf{p})\big]\;,
\end{equation}
and the free Hamiltonian $\hat{\bm{H}}^0$ can be written as
\begin{equation}
  \hat{\bm{H}}^0\ =\ \EP\hat{H}^0\:-\:\EM\hat{H}^{0*}\ =\ \hat{H}^0_+\:+\:\hat{H}^0_-\;,
\end{equation}
where
\begin{equation}
  \label{eq:Hamiltonian}
  \hat{H}_{\pm}^0 \ =\ \pm\,\frac{1}{2} \int\!{\rm d}^3\mathbf{x}\;\Big[\,\dot{\!\hat{\phi}}_{\pm}\!\!\!{}^2(t,\mathbf{x})\:
  +\: \big[\bm{\nabla}\hat{\phi}_{\pm}(t,\mathbf{x})\big]^2\:+\:m^2\hat{\phi}^2_{\pm}(t,\mathbf{x})\Big]\;.
\end{equation}
By virtue of the orthogonality of the idempotents, there can be no terms in the Hamiltonian that mix the positive- and negative-energy states. Moreover, since any product of positive and negative operators vanishes, operators that mix the positive- and negative-energy states also cannot be generated by radiative corrections.

The positive- and negative-energy, interaction-picture field operators have the familiar plane-wave decompositions
\begin{equation}
  \hat{\phi}_{\pm}(t,\mathbf{x})\ =\ \int\!{\rm d}\Pi_{\mathbf{p}}\;
  \Big[\hat{a}_{\pm}(0, \mathbf{p})\,e^{\mp\I E(\mathbf{p})x^0}{e}^{+\I \mathbf p \cdot \mathbf x}\:
  +\: \hat{a}^{\#}_{\pm}(0,\mathbf{p})\,e^{\pm\I E(\mathbf{p})x^0}e^{-\I \mathbf p \cdot \mathbf x} \Big]\;,
\end{equation}
and satisfy the equal-time commutation relations
\begin{subequations}
\begin{align}
  \big[\hat{\phi}_{\pm}(t,\mathbf{x}),\hat{\phi}_{\pm}(t,\mathbf{y})\big]\ &=\ \0\;,\\
  \label{eq:piphi}
  \big[\hat{\pi}_{\pm}(t,\mathbf{x}),\hat{\phi}_{\pm}(t,\mathbf{y})\big]\ &=\ \pm\, \I\EPM\delta^3(\mathbf{x}-\mathbf{y})\;.
\end{align}
\end{subequations}
The time-ordered propagators of the positive- and negative-energy fields are the Feynman and Dyson propagators, respectively:
\begin{subequations}
\label{eq:Deltas}
\begin{align}
  \langle 0 |T\hat{\phi}_{+}(x)\hat{\phi}_{+}(y)|0\rangle\ =\ \Delta_{\rm F}(x,y)\ 
  =\ +\,\I\,\EP\int\frac{{\rm d}^4p}{(2\pi)^4}\;\frac{e^{-ip\cdot(x-y)}}{p^2-m^2+i\epsilon}\;,\\
  \langle 0 |T\hat{\phi}_{-}(x)\hat{\phi}_{-}(y)|0\rangle\ =\ \Delta_{\rm D}(x,y)\
  =\ -\,\I\,\EM\int\frac{{\rm d}^4p}{(2\pi)^4}\;\frac{e^{-ip\cdot(x-y)}}{p^2-m^2-i\epsilon}\;,
\end{align}
\end{subequations}
where $\epsilon\to 0^+$.

Inserting the field operators into Eq.~\eqref{eq:Hamiltonian} and making use of the commutation relations of the creation and annihilation operators, the free Hamiltonian can be written in the form
\begin{equation}
  \label{eq:scalarH}
  \hat{\bm{H}}^0\ =\ \frac{1}{2} \int\! \frac{{\rm d}^3\mathbf{p}}{(2\pi)^3}\; \Big[\hat{a}_{+}^{\#}(0, \mathbf p)\hat{a}_{+}(0, \mathbf p)\:
  -\:  {\hat{a}_{-}}^{\#}(0, \mathbf p)\hat{a}_{-}(0, \mathbf p)\:+\:(2\pi)^3E(\mathbf{p})\delta^3(\mathbf{0})\K\Big]\;.
\end{equation}
By applying the results from Sec.~\ref{sec:algebra}, we can show straightforwardly that the vacuum expectation value of the Hamiltonian $\hat{\bm{H}}^0$ is zero. Most significantly, the zero-point energy cancels between the positive- and negative-energy components, since
\begin{equation}
  \llangle 0|\K|0\rrangle\ =\ \llangle 0|\EP|0\rrangle\:-\:\llangle 0|\EM|0\rrangle\ = \ \llangle 0_+|0_+\rrangle\:-\:\llangle 0_-|0_-\rrangle\ =\ 0\;.
\end{equation}
In this way, composite operators that are anti-symmetric in positive- and negative-energy components are effectively normal-ordered (cf.~Ref.~\cite{Cartas-Fuentevilla:2017ged}). By the same token, there is no zero-point contribution to the vacuum expectation value of the energy-momentum tensor. This, of course, presents an intriguing possibility for the resolution of the cosmological constant problem, and we will make some final remarks in the context of gravity in Sec.~\ref{sec:gravity}.

The cancellation of the zero-point energy has occurred at the level of the Euclidean inner product. Whilst the zero-point contribution is present in the Heisenberg equation, it does not contribute to the evolution of the operators or corresponding states, since the algebra of the bicomplex numbers is commutative, i.e. $\K\cdot \EPM\ =\ \EPM \cdot \K$. The zero-point contribution therefore trivially commutes with the creation and annihilation operators, as it does in the case of standard quantum field theory. We can then show that the positive- and negative-energy, single-particle Fock states are eigenstates of the normal-ordered Hamiltonian $:\hat{\bm{H}}^0:$, with eigenvalues $+E(\mathbf{p})$ and $-E(\mathbf{p})$, respectively, i.e.
\begin{equation}
  :\hat{\bm{H}}^0:|t,\mathbf{p}_{\pm}\rrangle\ =\ \pm\,E(\mathbf{p})|t,\mathbf{p}_{\pm}\rrangle\;.
\end{equation}
Hence, at the level of the free theory, we have managed to introduce positive- and negative-energy states, which cannot mix and have positive norms, and whose interplay leads to the cancellation of the zero-point energy.


\subsection{Discrete symmetry transformations}
\label{sec:discrete}

Along with the usual charge-conjugation $\rm C$, parity $\rm P$ and time-reversal $\rm T$ transformations, we can introduce an energy-parity transformation $\rm E$ that relates the positive- and negative-energy components. Specifically, we have the following set of discrete symmetry 
transformations:
\begin{subequations}
\begin{align}
  {\rm C}:&\qquad \hat{\mathcal{C}}\hat{\phi}_{\pm}(t,\mathbf{x})\hat{\mathcal{C}}^{-1}\ =\ \hat{\phi}_{\pm}(t,\mathbf{x})\;,\\
  {\rm P}:&\qquad \hat{\mathcal{P}}\hat{\phi}_{\pm}(t,\mathbf{x})\hat{\mathcal{P}}^{-1}\ =\ \hat{\phi}_{\pm}(t,-\,\mathbf{x})\;,\\
  {\rm T}:&\qquad \hat{\mathcal{T}}\hat{\phi}_{\pm}(t,\mathbf{x})\hat{\mathcal{T}}^{-1}\ =\ \hat{\phi}_{\pm}(-\,t,\mathbf{x})\;,\\
  {\rm E}:&\qquad \hat{\mathcal{E}}\hat{\phi}_{\pm}(t,\mathbf{x})\hat{\mathcal{E}}^{-1}\ =\ \hat{\phi}_{\mp}(t,\mathbf{x})\;.
\end{align}
\end{subequations}
Of particular interest are the time-reversal and energy-parity transformations. The transformations of the creation and annihilation operators are as follows:
\begin{subequations}
\begin{align}
  {\rm T}:&\qquad \hat{\mathcal{T}}\hat{a}_{\pm}(0,\mathbf{p})\hat{\mathcal{T}}^{-1}\ =\ \hat{a}_{\pm}(0,-\,\mathbf{p})\;,\\
  &\qquad \hat{\mathcal{T}}\hat{a}^{\#}_{\pm}(0,\mathbf{p})\hat{\mathcal{T}}^{-1}\ =\ \hat{a}^{\#}_{\pm}(0,-\,\mathbf{p})\;,\\
  {\rm E}:&\qquad \hat{\mathcal{E}}\hat{a}_{\pm}(0,\mathbf{p})\hat{\mathcal{E}}^{-1}\ =\ \hat{a}_{\mp}(0,-\,\mathbf{p})\;,\\
  &\qquad \hat{\mathcal{E}}\hat{a}^{\#}_{\pm}(0,\mathbf{p})\hat{\mathcal{E}}^{-1}\ =\ \hat{a}_{\mp}^{\#}(0,-\,\mathbf{p})\;.
\end{align}
\end{subequations}
Whereas $\mathcal{T}$ must be anti-unitary with respect to \emph{both} $\I$ and $\J$ (i.e.~$\forall\lambda\in\mathbb{BC}$, $\mathcal{T}\lambda\mathcal{T}^{-1}=\lambda^{\times}$), $\mathcal{E}$ must be anti-unitary with respect to $\I$ only (i.e.~$\forall\lambda\in\mathbb{BC}$, $\mathcal{E}\lambda\mathcal{E}^{-1}=\lambda^{*}$). We note that the Hamiltonian $\hat{\bm{H}}^0$ is even under $\rm T$ and odd under $\rm E$.


\section{Haag's theorem}
\label{sec:Haag}

In order to illustrate the deficiencies of the standard Fock representation, we follow Haag's original arguments~\cite{Haag:1955ev}. We consider two free scalar fields with masses $m_1$ and $m_2$, satisfying the Klein-Gordon equations
\begin{align}
  (\Box\:+\:m_1^2)\hat{\phi}_1(x)\ &=\ 0\;,\\
  (\Box\:+\:m_2^2)\hat{\phi}_2(x)\ &=\ 0\;,
\end{align}
and comprising \emph{only} positive-energy modes. If the fields $\hat{\phi}_1(x)$ and $\hat{\phi}_2(x)$, and their time-derivatives coincide at $t=0$, i.e.
\begin{align}
  \hat{\phi}_1(0,\mathbf{x})\ &=\ \hat{\phi}_2(0,\mathbf{x})\ =\ \hat{\phi}(0,\mathbf{x})\;,\\
  \,\dot{\!\hat{\phi}}_1(0,\mathbf{x})\ &=\ \,\dot{\!\hat{\phi}}_2(0,\mathbf{x})\ =\ \hat{\pi}(0,\mathbf{x})\;,
\end{align}
then we can write the creation and annihilation operators
\begin{align}
  \hat{a}_{\ell}(\mathbf{p})\ &=\ \int\!{\rm d}^3\mathbf{x}\;\Big(E_{\ell}(\mathbf{p})\hat{\phi}(0,\mathbf{x})\:
  +\:i\hat{\pi}(0,\mathbf{x})\Big)e^{-i\mathbf{p}\cdot \mathbf{x}}\;,\\
  \hat{a}^{\dag}_{\ell}(\mathbf{p})\ &=\ \int\!{\rm d}^3\mathbf{x}\;\Big(E_{\ell}(\mathbf{p})\hat{\phi}(0,\mathbf{x})\:
  -\:i\hat{\pi}(0,\mathbf{x})\Big)e^{+i\mathbf{p}\cdot \mathbf{x}}\;,
\end{align}
where $\ell=1,2$. The creation and annihilation operators of the two fields are therefore related by a Bogoliubov transformation of the form
\begin{equation}
  \hat{a}_2(\mathbf{p})\ =\ \frac{E_1+E_2}{2E_1}\,\hat{a}_1(\mathbf{p})\:-\:\frac{E_1-E_2}{2E_2}\,\hat{a}_1^{\dag}(\mathbf{p})\;.
\end{equation}
Herein, we have omitted the three-momentum arguments of the energies $E_1$ and $E_2$, and suppressed the time-dependence of the operators for notational convenience. It immediately follows that there does \emph{not} exist a vacuum state $|0\rangle$ for which $\hat{a}_1(\mathbf{p})|0\rangle$ and $\hat{a}_2(\mathbf{p})|0\rangle$ are simultaneously zero, except when $m_1=m_2$. Suppose we choose the vacuum state to be annihilated by $\hat{a}_1(\mathbf{p})$. The number operator
\begin{align}
  \hat{n}_2(\mathbf{p})\ &=\ \frac{1}{V}\,\frac{1}{2E_2}\,\hat{a}^{\dag}_2(\mathbf{p})\hat{a}_2(\mathbf{p})\nonumber\\ &
  =\ \frac{1}{V}\,\frac{E_1^2+E_2^2}{4E_1^2E_2}\,\hat{a}_1^{\dag}(\mathbf{p})\hat{a}_1(\mathbf{p})\:
  +\: \frac{1}{V}\,\frac{E_1^2-E_2^2}{8E_1^2E_2}\Big(\hat{a}_1^{\dag}(\mathbf{p})\hat{a}^{\dag}_1(\mathbf{p})
  +\hat{a}_1(\mathbf{p})\hat{a}_1(\mathbf{p})\Big)\nonumber\\&\qquad+\:\frac{\big(E_1-E_2\big)^2}{4E_1E_2}
\end{align}
will have a non-zero vacuum expectation value:
\begin{equation}
  \langle 0| \hat{n}_2(\mathbf{p})|0 \rangle\ =\ \frac{\big(E_1-E_2\big)^2}{4E_1E_2}\;.
\end{equation}

We now repeat the same arguments for the bicomplex extension of this theory. The relevant number operator is
\begin{equation}
  \hat{\bm{n}}(\mathbf{p})\ \equiv\ \hat{a}_{+}^{\#}(\mathbf{p})\hat{a}_{+}(\mathbf{p})\:
  -\:\hat{a}_{-}^{\#}(\mathbf{p})\hat{a}_{-}(\mathbf{p})\;,
\end{equation}
the definition of which, including the relative sign between the positive- and negative-energy components, follows by inspection of the Hamiltonian in Eq.~\eqref{eq:scalarH}. For the two-field model above, we therefore have
\begin{align}
  \hat{\bm{n}}_2(\mathbf{p})\ &=\ \frac{E_1^2+E_2^2}{2E_1^2}\,\hat{\bm{n}}_1(\mathbf{p})\nonumber\\&
  +\ \frac{1}{V}\,\frac{E_1^2-E_2^2}{8E_1^2E_2}\sum_{\pm}\pm\Big(\hat{a}_{1\pm}^{\dag}(\mathbf{p})\hat{a}^{\dag}_{1\pm}(\mathbf{p})
  +\hat{a}_{1\pm}(\mathbf{p})\hat{a}_{1\pm}(\mathbf{p})\Big)\nonumber\\&+\ \frac{\big(E_1-E_2\big)^2}{4E_1E_2}\,\K\;.
\end{align}
In contrast to the standard case, there now \emph{does} exist a unique vacuum state for which the vacuum expectation values of both $\hat{\bm{n}}_1$ and $\hat{\bm{n}}_2$ vanish simultaneously, i.e.
\begin{equation}
  \llangle 0|\hat{\bm{n}}_1(\mathbf{p})|0\rrangle \ = \ 0\;,\qquad \llangle 0|\hat{\bm{n}}_2(\mathbf{p})|0\rrangle\ =\ 0\;.
\end{equation} 
This result suggests that the present construction may circumvent Haag's theorem and the associated criticism of the Fock representation. A proof that this is the case would require us to show that, in contrast to the standard unitary evolution operator, the bicomplex evolution operator $\hat{\bm{U}}$ is a \emph{proper} unitary operator, and this may be presented elsewhere.


\section{Interacting theory}
\label{sec:inttheory}

As a playground in which to study interactions in this construction, we consider the archetypal $\phi^4$ theory. The bicomplex extension of its interaction Hamiltonian is not unique, since the hyperbolic unit $\K=\K^{\#}$ enables us to construct a number of interaction Hamiltonians $\hat{\bm{H}}^{\rm int}(t)$, all of which yield an evolution operator
\begin{equation}
  \hat{\bm{U}}(t,t')\ =\ T\,\exp\bigg[-\,i\int_{t'}^{t}{\rm d}t''\;\hat{\bm{H}}^{\rm int}(t'')\bigg]
\end{equation}
that is unitary with respect to the bicomplex adjoint, i.e.~$\hat{\bm{U}}^{\#}(t,t')\hat{\bm{U}}(t,t') = \1\otimes\hat{\mathbb{I}}$. In the present work, we will consider two examples. In the first, the interaction Hamiltonian will be chosen so as to reflect the same energy-parity symmetry as the free part of the Hamiltonian. This choice will ensure that contributions to the vacuum energy continue to cancel at the loop level, and we will show that there exists a non-trivial scattering matrix. In the second case, we will find results entirely equivalent to Ref.~\cite{Dickinson:2015ixa}, yielding tree-level matrix elements consistent with standard results and reduced sensitivity to ultraviolet divergences, but giving rise to vacuum energy corrections at the two-loop level.


\subsection{Energy-parity asymmetric}
\label{sec:asymmetric}

We first consider an interaction Hamiltonian that is odd under the energy-parity transformation defined in Sec.~\ref{sec:discrete}:
\begin{equation}
  \bm{\hat{H}}^{\rm int}(t)\ =\ \int{\rm d}^3\mathbf{x}\;\Bigg[\frac{\lambda}{4!}\,\hat{\phi}_+^4(t,\mathbf{x})\:
  -\:\frac{\lambda}{4!}\,\hat{\phi}_-^4(t,\mathbf{x})\Bigg]\;.
\end{equation}
Having shown already that the $\llangle 0 |\hat{\bm{H}}^0|0\rrangle=0$, the vacuum expectation value of the full Hamiltonian $\hat{\bm{H}}(t)=\hat{\bm{H}}^0\:+\:\hat{\bm{H}}^{\rm int}(t)$ reduces to
\begin{align}
  \label{eq:vacs1}
  \llangle 0|\hat{\bm{H}}(t)|0\rrangle\ &=\ \frac{\lambda}{4!}\int{\rm d}^3\mathbf{x}\;\llangle 0|\big[\hat{\phi}_+^4(t,\mathbf{x})\:
  -\:\hat{\phi}_-^4(t,\mathbf{x})\big]|0\rrangle\nonumber\\
  &=\ \frac{\lambda}{16}\,V_3\,\Big[\Delta_{\rm F}^2(0)\:-\:\Delta_{\rm D}^2(0)\Big]\;,
\end{align}
where $V_3$ is a three-volume factor. Since
\begin{equation}
  \Delta_{\rm F}(0)\ =\ \Delta_{\rm D}(0)\ \in \ \mathbb{R}\;,
\end{equation}
[see Eq.~\eqref{eq:Deltas}] we immediately see that the leading loop corrections to the vacuum energy also cancel, and this proceeds to all orders.

In order to generate the $n$-point functions of this theory, we introduce an external and real-valued source $J(x)$, whose couplings to the positive- and negative-energy modes are also odd under ${\rm E}$:
\begin{equation}
  \label{eq:HamJ1}
  \bm{\hat{H}}^{\rm int}(t)\ =\ \int{\rm d}^3\mathbf{x}\;\Bigg[\frac{\lambda}{4!}\,\hat{\phi}_+^4(t,\mathbf{x})\:
  -\:\frac{\lambda}{4!}\,\hat{\phi}_-^4(t,\mathbf{x})\:-\:J(t,\mathbf{x})\hat{\phi}_+(t,\mathbf{x})\:
  +\:J(t,\mathbf{x})\hat{\phi}_-(t,\mathbf{x})\Bigg]\;.
\end{equation}
We emphasise that requiring the full Hamiltonian to be odd under ${\rm E}$, the classical sources interact with both the positive- and negative-energy degrees of freedom. Whilst direct mixing of the positive- and negative-energy states cannot occur, the coupling of classical sources to both is not precluded. The $n$-point functions are then generated from the vacuum persistence amplitude
\begin{equation}
  \mathcal{Z}[J]\ =\ \llangle \Omega(+\infty)|\Omega(-\infty)\rrangle_J\ =\ \llangle 0|\hat{\bm{U}}(+\infty,-\infty)|0\rrangle
\end{equation}
through functional differentiation with respect to $J(x)$:
\begin{equation}
G_n(x_1,x_2,\dots,x_n)\ =\ \frac{1}{\mathcal{Z}[0]}\,\Bigg[\prod_{\ell\,=\,1}^n\frac{1}{i}\,\frac{\delta}{\delta J(x_{\ell})}\Bigg]\mathcal{Z}[J]\Bigg|_{J\,=\,0}\;.
\end{equation}
In the absence of the source $J(x)$, the vacuum persistence amplitude is unity, since all vacuum loops cancel, i.e.
\begin{equation}
  \mathcal{Z}[0]\ =\ 1\;.
\end{equation}
We might expect therefore that we will end up with a trivial theory.

The corresponding scattering matrix theory can be constructed from asymptotic states of the form
\begin{equation}
  \label{eq:asymp}
  |\mathbf{p},{\rm in}\rrangle\ =\ |\mathbf{p}_+,{\rm in}\rrangle\:-\:|\mathbf{p}_-,{\rm in}\rrangle\ =\ Z_{\phi}^{-1/2}\Big(|0,\mathbf{p}_+\rrangle\:-\:|0,\mathbf{p}_-\rrangle\Big)\;,
\end{equation}
with the scattering operator
\begin{equation}
  \hat{\bm{S}}\ =\ \hat{\bm{U}}(+\,\infty,-\,\infty)\;.
\end{equation}
Hereafter, we will suppress factors of the wavefunction renormalization $Z_{\phi}$. Noticing that the asymptotic states are \emph{not} eigenstates of the free Hamiltonian $\hat{\bm{H}}^0$, and that
\begin{equation}
  \llangle \mathbf{p},{\rm in}|\hat{\bm{H}^0}|\mathbf{p},{\rm in}\rrangle\ =\ 0\;,
\end{equation}
we might again be led to believe that we are dealing with a trivial theory.

By taking the second variation of $\mathcal{Z}[J]$, we find that the leading contribution to the $2$-point function is
\begin{equation}
  G_2^{(0)}(x_1,x_2)\ =\ \frac{1}{2}\Big[\Delta_{\rm F}(x_1,x_2)\:+\:\Delta_{\rm D}(x_1,x_2)\Big]\
  =\ \int\frac{{\rm d}^4p}{(2\pi)^4}\,\pi\delta(p^2-m^2)e^{-ip\cdot(x_1-x_2)}\;.
\end{equation}
This is problematic, since only the on-shell part of the propagator has survived. A similar situation arises for the $4$-point function:
\begin{equation}
  G_4^{(0)}(x_1,x_2,x_3,x_4)\ =\ \frac{-\,i\lambda}{2}\int{\rm d}^4z\Bigg[\prod_{\ell\,=\,1}^4\Delta_{\rm F}(x_{\ell},z)\:
  -\:\prod_{\ell\,=\,1}^4\Delta_{\rm D}(x_{\ell},z)\Bigg]\;.
\end{equation}
We can effect the LSZ reduction~\cite{Lehmann:1954rq} and amputate the external legs by acting on the $n$-point function with
\begin{equation}
  \prod_{\ell\,=\,1}^4\int{\rm d}^4x_{\ell}\;e^{ip_{\ell}\cdot x_{\ell}}\,i\big(\Box^2+m^2\big)\;,\qquad
  p_{\ell}^0\ =\ E(\mathbf{p}_{\ell})\;,
\end{equation}
where we have suppressed factors of the wavefunction renormalization $Z_{\phi}$.  Doing so, we find that the interplay of the positive- and negative-energy modes means that there are always an odd number of on-shell legs, and therefore
\begin{equation}
  \label{eq:vanishgamma4}
  \Gamma_4^{(0)}(p_1,p_2,p_3,p_4)\ =\ \Bigg[\prod_{\ell\,=\,1}^4\int{\rm d}^4x_{\ell}\;e^{ip_{\ell}\cdot x_{\ell}}\,i\big(\Box^2+m^2\big)\Bigg]G_4^{(0)}(x_1,x_2,x_3,x_4)\
  =\ 0\;.
\end{equation}
The result in Eq.~\eqref{eq:vanishgamma4} would appear to confirm our expectation of a trivial theory. As we will now show, however, the theory remains non-trivial.

Since $\mathcal{Z}[0]=1$, there is no distinction between the generating functional of connected and disconnected Green's function. The corollary is that disconnected diagrams involving vacuum bubbles cannot be cancelled by dividing through by $\mathcal{Z}[0]$, as they are in standard quantum field theory. As we will see, this peculiarity marks a breakdown of naive perturbation theory, and it may turn out to be pivotal for the viability of the interacting theory. Incidentally, we would still find that the disconnected diagrams would not cancel if we were to restrict only to positive-energy external states, i.e.~if we were to couple the external source only to the positive-energy field $\hat{\phi}_+(x)$, explicitly breaking the energy-parity symmetry.

Returning to the $2$-point function, at the next order, the disconnected diagram is
\begin{equation}
  G_2^{(1)}(x_1,x_2)\ \supset\ \frac{-\,i\lambda}{16}\,V_4\,\Big[\Delta_{\rm F}(x_1,x_2)\Delta_{\rm F}^2(0)\:
  -\:\Delta_{\rm D}(x_1,x_2)\Delta_{\rm D}^2(0)\Big]\;,
\end{equation}
where $V_4$ is a four-volume factor. Since $\Delta_{\rm F}(0)=\Delta_{\rm D}(0)$ [see Eq.~\eqref{eq:Deltas}], we find
\begin{equation}
  G_2^{(1)}(x_1,x_2)\ \supset\ i\,\mathcal{Z}_{\K}[0]\int\!\frac{{\rm d}^4p}{(2\pi)^4}\;{\rm PV}\,\frac{e^{-ip\cdot(x_1-x_2)}}{p^2-m^2}\;,
\end{equation}
where ${\rm PV}$ indicates the Cauchy principal value and
\begin{equation}
  \mathcal{Z}_{\K}[0]\ =\ \llangle \Omega(+\infty)|\K|\Omega(-\infty) \rrangle_{J\,=\,0}\
  =\ \frac{-i\lambda}{8}\,V_4\,\Delta_{\rm F}^2(0)\:+\:\mathcal{O}(\lambda^2)
\end{equation}
contains the vacuum loops. In the case of the $4$-point function, the next order in perturbation theory yields
\begin{align}
  G_4^{(1)}(x_1,x_2,x_3,x_4)\ &=\ \frac{(-i\lambda)^2}{2!(4!)^2}\int{\rm d}^4z_1\int{\rm d}^4z_2\;
  \llangle 0|T\Big[\hat{\phi}_+(x_1)\hat{\phi}_+(x_2)\hat{\phi}_+(x_3)\hat{\phi}_+(x_4)\hat{\phi}_+^4(z_1)\hat{\phi}_+^4(z_2)
  \nonumber\\&\qquad
  +\:\hat{\phi}_-(x_1)\hat{\phi}_-(x_2)\hat{\phi}_-(x_3)\hat{\phi}_-(x_4)\hat{\phi}_-^4(z_1)\hat{\phi}_-^4(z_2)\:\Big]|0\rrangle\;.
\end{align}
The disconnected diagram is
\begin{equation}
  G_4^{(1)}(x_1,x_2,x_3,x_4)\ \supset\ \frac{(-\,i\lambda)}{2}\,Z_{\K}[0]\int\!{\rm d}^4z\;
  \Bigg[\prod_{\ell\,=\,1}^4\Delta_{\rm F}(x_{\ell},z)\:+\:\prod_{\ell\,=\,1}^4\Delta_{\rm D}(x_{\ell},z)\Bigg]\;.
\end{equation}
On amputating the external legs, we now obtain
\begin{equation}
  \Gamma_4(p_1,p_2,p_3,p_4)\ \supset\ (-i\lambda)\mathcal{Z}_{\K}[0](2\pi)^4\delta^4\big({\textstyle \sum_{\ell\,=\,1}^4}p_{\ell}\big)\;,
\end{equation}
such that we appear to have recovered a non-zero matrix element, albeit one that is proportional to a formally divergent quantity.

The connected contribution to $G_4^{(1)}(x_1,x_2,x_3,x_4)$ is most easily expressed in momentum space. It is given by
\begin{align}
   G^{(1)}_4(p_1,p_2,p_3,p_4)\ &\supset\ \frac{1}{2}\sum_{q^2\,=\,s,\,t,\,u}\Bigg[\prod_{\ell\,=\,1}^4\Delta_{\rm F}(p_{\ell})i\Pi_{\lambda}(q^2)\:
  -\:\prod_{\ell\,=\,1}^4\Delta_{\rm D}(p_{\ell})i\Pi_{\lambda}^*(q^2)\Bigg]\nonumber\\&\qquad\times\:(2\pi)^4\delta^4\big({\textstyle \sum_{\ell\,=\,1}^4}p_{\ell}\big)\;,
\end{align}
where $s$, $t$ and $u$ are the Mandelstam variables and
\begin{equation}
  \label{eq:Pi}
  i\Pi_{\lambda}(q^2)\ =\ \frac{(-\,i\lambda)^2}{2}\int\!\frac{{\rm d}^4k}{(2\pi)^4}\;\frac{i}{k^2-m^2+i\epsilon}\,\frac{i}{(q-k)^2-m^2+i\epsilon}
\end{equation}
is the usual bubble diagram. Putting everything together, and on amputating the external legs, we find
\begin{equation}
  \Gamma_4(p_1,p_2,p_3,p_4)\ \supset\ \Bigg[(-i\lambda)\mathcal{Z}_{\K}[0]\:-\:\sum_{q^2\,=\,s,\,t,\,u}{\rm Im}\,\Pi_{\lambda}(q^2)\Bigg]
  (2\pi)^4\delta^4\big({\textstyle \sum_{\ell\,=\,1}^4}p_{\ell}\big)\;,
\end{equation}
where only the absorptive part of the one-loop corrections appears. Thus, at this order in naive perturbation theory, it would seem that we are not sensitive to the ultraviolet divergences in the real part of $\Pi_{\lambda}(q^2)$. However, these dispersive corrections reappear at third order, as a result of the interplay with the vacuum fluctuations. Specifically, we find
\begin{align}
  \label{eq:gamma41}
  \Gamma_4(p_1,p_2,p_3,p_4)\ &\supset\ \Bigg[(-i\lambda)\mathcal{Z}_{\K}[0]\:-\:\sum_{q^2\,=\,s,\,t,\,u}{\rm Im}\,\Pi_{\lambda}(q^2)\:
  +\:i\mathcal{Z}_{\K}[0]\sum_{q^2\,=\,s,\,t,\,u}{\rm Re}\,\Pi_{\lambda}(q^2)\Bigg]\nonumber\\&\qquad\times\:
  (2\pi)^4\delta^4\big({\textstyle \sum_{\ell\,=\,1}^4}p_{\ell}\big)\;.
\end{align}

Since $i\mathcal{Z}_{\K}[0]\in\mathbb{R}$, the matrix elements of this construction are real, and we should be concerned that we will lose interference effects. Such interference effects are, of course, pivotal in the case of gauge theories, wherein the removal of infrared divergences relies on cancellations between tree-level and one-loop diagrams by means of the Bloch-Nordsieck~\cite{Bloch:1937pw}, Kinoshita-Lee-Nauenberg~\cite{Kinoshita:1962ur,Lee:1964is} or Weinberg soft-graviton~\cite{Weinberg:1965nx} theorems. It is interesting, however, to consider the modulus squared of the four-point amplitude at order $\lambda^3$:
\begin{align}
  \label{eq:Gamma4sq}
  |\Gamma_4(p_2,p_2,p_3,p_4)|^2\ &=\ \Bigg[\Bigg(\lambda^2\:-\:2\lambda \sum_{q^2\,=\,s,\,t,\,u}{\rm Re}\,\Pi_{\lambda}(q^2)\Bigg)|\mathcal{Z}_{\K}[0]|^2\nonumber\\&\qquad
  +\:2\lambda |\mathcal{Z}_{\K}[0]|\sum_{q^2\,=\,s,\,t,\,u}{\rm Im}\,\Pi_{\lambda}(q^2)\Bigg]
  V_4(2\pi)^4\delta^4\big({\textstyle \sum_{\ell\,=\,1}^4}p_{\ell}\big)\;.
\end{align}
This is to be compared with the standard result
\begin{equation}
  |\Gamma_4(p_1,p_2,p_3,p_4)|^2\ =\ \Bigg[\lambda^2\:-\:2\lambda\sum_{q^2\,=\,s,\,t,\,u} {\rm Re}\,\Pi_{\lambda}(q^2)\Bigg]
  V_4(2\pi)^4\delta^4\big({\textstyle \sum_{\ell\,=\,1}^4}p_{\ell}\big)\:+\:\mathcal{O}(\lambda^4)\;.
\end{equation}
The presence of the absorptive correction at order $\lambda^3$ in Eq.~\eqref{eq:Gamma4sq} would appear to signal potentially serious problems with perturbative unitarity. However, the absorptive correction is suppressed relative to the tree-level and dispersive corrections by a factor of $|\mathcal{Z}_{\K}[0]|$, such that one might infer that it is subleading. Whilst this is an interesting possibility, it is clear that naive perturbation theory may be misleading for the present construction. Not least of all, a careful treatment of its renormalization is needed, wherein the resummation of factors of $|\mathcal{Z}_{\K}[0]|$ quickly becomes non-trivial. We leave a dedicated study of these issues for future work. Even so, the restoration of perturbative unitarity, and any consistency of transition probabilities with standard results, will certainly rely on this peculiar interplay of the vacuum fluctuations.

The results presented above carry over to the $\phi^3$ theory with interaction Hamiltonian
\begin{equation}
  \hat{\bm{H}}^{\rm int}(t)\ =\ \int{\rm d}^3\mathbf{x}\;\Bigg[\frac{\kappa}{3!}\,\hat{\phi}^3_+(t,\mathbf{x})\:
  -\:\frac{\kappa}{3!}\,\hat{\phi}^3_-(t,\mathbf{x})\:-\:J(t,\mathbf{x})\hat{\phi}_+(t,\mathbf{x})\:
  +\:J(t,\mathbf{x})\hat{\phi}_-(t,\mathbf{x})\Bigg]\;.
\end{equation}
For example, the amplitude for the two-to-two scattering process is zero at leading order, and it again reappears only through the interplay with the vacuum fluctuations, i.e.
\begin{equation}
  \Gamma_4(p_1,p_2,p_3,p_4)\ \supset\ (-i\kappa)^2i\mathcal{Z}_{\K}[0]\sum_{q^2\,=\,s,\,t,\,u}{\rm PV}\,\frac{1}{q^2-m^2}\,
  (2\pi)^4\delta^4\big({\textstyle \sum_{\ell\,=\,1}^4}p_{\ell}\big)\;,
\end{equation}
where
\begin{equation}
  \mathcal{Z}_{\K}[0]\ =\ i(-i\kappa)^2\,{\rm Im}\int\!{\rm d}^4x\int\!{\rm d}^4y\Bigg[\frac{1}{8}\,\Delta_{\rm F}^2(0)\Delta_{\rm F}(x,y)\:
  +\:\frac{1}{12}\,\Delta_{\rm F}^3(x,y)\Bigg]\:+\:\dots\;.
\end{equation}

We remark that there exists an alternative definition of the scattering operator: $\hat{\bm{S}}\to \K\hat{\bm{S}}$, which is still unitary with respect to the bicomplex adjoint. For this choice, we might conclude that we obtain a trivial agreement with the usual tree-level matrix elements. However, the interplay of the vacuum fluctuations remains non-trivial. For example, the leading contribution to the two-point function, which is now obtained from
\begin{align}
  \label{eq:Gk2def}
  G_{\K2}(x_1,x_2)\ &=\ \frac{1}{\mathcal{Z}_{\K}[0]}\,\frac{1}{i}\,\frac{\delta}{\delta J(x_1)}\,\frac{1}{i}\,\frac{\delta}{\delta J(x_2)}\,
  \mathcal{Z}_{\K}[J]\bigg|_{J\,=\,0}\;,
\end{align}
becomes
\begin{align}
  G_{\K2}^{(0)}(x_1,x_2)\ &=\ \frac{1}{2}\,\frac{1}{\mathcal{Z}_{\K}[0]}\Big[\Delta_{\rm F}(x_1,x_2)\:-\:\Delta_{\rm D}(x_1,x_2)\Big]
  \nonumber\\ &=\ \frac{i}{\mathcal{Z}_{\K}[0]}\int\!\frac{{\rm d}^4p}{(2\pi)^4}\,{\rm PV}\,\frac{e^{-ip\cdot (x_1-x_2)}}{p^2-m^2}\;,
\end{align}
where the inverse of $\mathcal{Z}_{\K}[0]$ has arisen from the normalization in Eq.~\eqref{eq:Gk2def}. The four-point amplitude of the $\phi^4$ theory becomes
\begin{align}
  \Gamma_{\K4}(p_2,p_2,p_3,p_4)\ &\supset\ \Bigg[(-i\lambda)\mathcal{Z}^{-1}_{\K}[0]\:-\:\sum_{q^2\,=\,s,\,t,\,u}{\rm Im}\,\Pi_{\lambda}(q^2)\:
  +\:i\mathcal{Z}^{-1}_{\K}[0]\sum_{q^2\,=\,s,\,t,\,u}{\rm Re}\,\Pi_{\lambda}(q^2)\Bigg]\nonumber\\&\qquad\times\:
  (2\pi)^4\delta^4\big({\textstyle \sum_{\ell\,=\,1}^4}p_{\ell}\big)\;,
\end{align}
resembling Eq.~\eqref{eq:gamma41} but differing in the appearance of factors of $\mathcal{Z}_{\K}[0]$.


\subsection{Energy-parity symmetric}
\label{sec:symmetric}

An energy-parity symmetric choice for the interaction Hamiltonian is:
\begin{equation}
  \bm{\hat{H}}^{\rm int}(t)\ =\ \int{\rm d}^3\mathbf{x}\;\Bigg[\frac{\lambda}{4!}\,\hat{\phi}_+^4(t,\mathbf{x})\:
  +\:\frac{\lambda}{4!}\,\hat{\phi}_-^4(t,\mathbf{x})\:-\:\K^{1/2}J(t,\mathbf{x})\hat{\phi}_+(t,\mathbf{x})\:
  +\:\K^{1/2}J(t,\mathbf{x})\hat{\phi}_-(t,\mathbf{x})\Bigg]\;,
\end{equation}
where the $\mathbb{C}_2(i)$ linear vector space representation of $\K^{1/2}$ is given in Eq.~\eqref{eq:rootk}. By absorbing the factors of $\K^{1/2}$ into the definitions of the interaction-picture operators via $\K^{1/2} a_{\pm}^{(\#)}(t,\mathbf{p})\ \to\ a_{\pm}^{(\#)}(t,\mathbf{p})$ and $\pm\,\K^{1/2}\hat{\phi}_{\pm}(x)\to \hat{\phi}_{\pm}(x)$, we recover precisely the construction in Ref.~\cite{Dickinson:2015ixa}, with the exception that direct mixing of the positive- and negative-energy components is precluded by the structure of the algebra.

Whilst the zero-point energy still cancels in the vacuum expectation value of $\hat{\bm{H}}^0$, it is clear that the two-loop contributions to the vacuum energy cannot cancel by straightforward comparison with Eq.~\eqref{eq:vacs1}. Moreover, these vacuum contributions will not cancel in the vacuum persistence amplitude. The leading contribution to the two-point function now takes the form
\begin{equation}
  G_2^{(0)}(x_1,x_2)\ =\ \frac{1}{2}\,\Big[\Delta_{\rm F}(x_1,x_2)\:-\:\Delta_{\rm D}(x_2,x_2)\Big]\
  =\ i\int\!\frac{{\rm d}^4p}{(2\pi)^4}\,{\rm PV}\,\frac{e^{-ip\cdot(x_1-x_2)}}{p^2-m^2}\;,
\end{equation}
wherein the principal value contribution has survived. The leading contribution to the four-point function is
\begin{equation}
  G_4^{(0)}(x_1,x_2,x_3,x_4)\ =\ \frac{-\,i\lambda}{2}\int{\rm d}^4z\Bigg[\prod_{\ell\,=\,1}^4\Delta_{\rm F}(x_{\ell},z)\:
  +\:\prod_{\ell\,=\,1}^4\Delta_{\rm D}(x_{\ell},z)\Bigg]\;,
\end{equation}
and this yields the usual amputated four-point amplitude
\begin{equation}
  \Gamma_4(p_1,p_2,p_3,p_4)\ =\ (-\,i\lambda)(2\pi)^4\delta^4\big({\textstyle \sum_{\ell\,=\,1}^4p_{\ell}}\big)\:+\:\mathcal{O}(\lambda^2)\;.
\end{equation}
Moving again to momentum space, the one-loop corrections to the four-point function are
\begin{align}
  G^{(1)}_4(p_1,p_2,p_3,p_4)\ &=\ \frac{1}{2}\sum_{q^2\,=\,s,\,t,\,u}\Bigg[\prod_{\ell\,=\,1}^4\Delta_{\rm F}(p_{\ell})i\Pi_{\lambda}(q^2)\:
  -\:\prod_{\ell\,=\,1}^4\Delta_{\rm D}(p_{\ell})i\Pi_{\lambda}^*(q^2)\Bigg]\nonumber\\&\qquad\times\:(2\pi)^4\delta^4\big({\textstyle \sum_{\ell\,=\,1}^4}p_{\ell}\big)\;,
\end{align}
and we find
\begin{equation}
  \Gamma_4(p_1,p_2,p_3,p_4)\ =\ \Bigg[(-i\lambda)\:-\:\sum_{q^2\,=\,s,\,t,\,u}{\rm Im}\,\Pi_{\lambda}(q^2)\Bigg]
  (2\pi)^4\delta^4\big({\textstyle \sum_{\ell\,=\,1}^4p_{\ell}}\big)\:+\:\mathcal{O}(\lambda^3)\;.
\end{equation}
In contrast to the previous case [Sec.~\ref{sec:asymmetric}], the disconnected diagrams now cancel with the normalization of the vacuum persistence amplitude. The interplay with the vacuum fluctuations therefore does not lead to a reappearance of the dispersive one-loop correction, viz.~${\rm Re}\,\Pi_{\lambda}(q^2)$, and, as observed in Ref.~\cite{Dickinson:2015ixa}, the one-loop result is ultraviolet finite. 

Whilst the above results may present an intriguing possibility for the hierarchy problem and naturalness, we find that, for this choice of interaction Hamitonian, all tree-level matrix elements are purely imaginary, all one-loop matrix elements are purely real, and so on. As a result, we should again be concerned about problems with perturbative unitarity. We note, however, that at order $\lambda^3$, the absorptive one-loop corrections still cancel in the modulus squared of the four-point amplitude:
\begin{equation}
  |\Gamma_4(p_1,p_2,p_3,p_4)|^2\ =\ \lambda^2\,V_4(2\pi)^4\delta^4\big({\textstyle \sum_{\ell\,=\,1}^4}p_{\ell}\big)\:+\:\mathcal{O}(\lambda^4)\;,
\end{equation}
such that this differs from the standard result only by the absence of the dispersive one-loop corrections.

The cubic interaction Hamiltonian consistent with Ref.~\cite{Dickinson:2015ixa} is
\begin{equation}
  \label{eq:HJ3}
  \bm{\hat{H}}^{\rm int}(t)\ =\ \K^{1/2}\int{\rm d}^3\mathbf{x}\;\Bigg[\frac{\kappa}{3!}\,\hat{\phi}_+^3(t,\mathbf{x})\:
  +\:\frac{\kappa}{3!}\,\hat{\phi}_-^3(t,\mathbf{x})\:-\:J(t,\mathbf{x})\hat{\phi}_+(t,\mathbf{x})\:
  +\:J(t,\mathbf{x})\hat{\phi}_-(t,\mathbf{x})\Bigg]\;,
\end{equation}
which can again be mapped directly onto the cubic theory in Ref.~\cite{Dickinson:2015ixa} by absorbing factors of $\K^{1/2}$ into the operators, as detailed above for the $\phi^4$ theory. The leading contribution to the three-point function is
\begin{equation}
  G_3^{(0)}(x_1,x_2,x_3)\ = \ \frac{-\,i\lambda}{2}\int\!{\rm d}^4z\Bigg[\prod_{\ell\,=\,1}^3\Delta_{\rm F}(x_{\ell},z)\:
  -\:\prod_{\ell\,=\,1}^3\Delta_{\rm D}(x_{\ell},z)\Bigg]\;,
\end{equation}
yielding the three-point amplitude
\begin{equation}
  \Gamma_3(p_1,p_2,p_3)\ =\ (-\,i\kappa)(2\pi)^4\delta\big({\textstyle \sum_{\ell\,=\,1}^3}p_{\ell}\big)\:+\:\mathcal{O}(\kappa^3)\;,
\end{equation}
which, although consistent with the usual result, of course vanishes by four-momentum conservation. The leading contribution to the four-point function is
\begin{align}
  \label{eq:2tot2phi3}
  \Gamma_4(p_1,p_2,p_3,p_4)\ = \ (-\,i\kappa)^2\,(2\pi)^4\delta^4\big({\textstyle \sum_{p_{\ell}\,=\,1}^4}p_{\ell}\big)\sum_{q^2\,=\,s,\,t,\,u}
  {\rm PV}\,\frac{i}{q^2-m^2}\:+\:\mathcal{O}(\kappa^4)\;,
\end{align}
which is again consistent with the usual result.

Proceeding beyond tree-level, the one-loop corrections to the two-point function take the following form in momentum space:
\begin{equation}
  \label{eq:G2phi3}
  G^{(1)}_2(p)\ =\ -\:{\rm PV}\,\bigg(\frac{1}{p^2-m^2}\bigg)^{\!2}\,{\rm Im}\,\Pi_{\kappa}(p^2)\:+\:\pi\delta'(p^2-m^2)\,{\rm Re}\,\Pi_{\kappa}(p^2)\;,
\end{equation}
where the self-energy $\Pi_{\kappa}(p^2)$ is given by Eq.~\eqref{eq:Pi}. In order to arrive at Eq.~\eqref{eq:G2phi3}, we have made use of the identity
\begin{equation}
  \Bigg(\frac{1}{p^2-m^2\pm i\epsilon}\Bigg)^{\!2}\ =\ {\rm PV}\,\Bigg(\frac{1}{p^2-m^2}\Bigg)^{\!2}\:\mp\:\pi\delta'(p^2-m^2)\;,
\end{equation}
in which $\delta'(p^2-m^2)=\partial_{p_0^2}\delta(p^2-m^2)$ is the derivative Delta function.  Inserting the one-loop result from Eq.~\eqref{eq:G2phi3} into the two-to-two scattering in Eq.~\eqref{eq:2tot2phi3}, wherein only the principal value contributes, we see the same reduced sensitivity to ultraviolet divergences (at this order) as for the $\phi^4$ theory.


\section{A note on negative-energy cascades and gravity}
\label{sec:gravity}

By virtue of the complementarity of the zero divisors (i.e.~$\EP\cdot\EM=\0$), operators that directly mix the positive- and negative-energy states cannot be generated through radiative effects, and one might anticipate that this structure is preserved when coupled to gravity. Dangerous negative-energy cascades would then be eliminated, and the vacuum of our theory would be stable, so long as the positive- and negative-energy sectors are stable in their own rights. Even so, we remark that it is possible to make sense of unstable potentials in the framework of $\mathcal{PT}$-symmetric quantum mechanics, see e.g.~Refs.~\cite{Bender:2005tb,Bender:2007nj}.

With such extensions to gravity in mind, we note that for a bicomplex matrix $\bm{A}\equiv  \EP\cdot A_+\:+\:\EM\cdot A_-$, where $A_{\pm}\in \mathbb{C}^n(\I)$, the $p$-th power of its determinant over $\mathbb{C}^n$ satisfies
\begin{equation}
  (\mathrm{det}_{\mathbb{C}^n(\I)}\,\mathbf{A})^{p}\ =\ \EP\cdot(\mathrm{det}\,A_+)^p\:+\:\EM\cdot(\mathrm{det}\,A_-)^p\;.
\end{equation}
It is therefore tempting to write the metric of general relativity as
\begin{equation}
  \bm{g}_{\mu\nu}\ =\ \1\cdot g_{\mu\nu}\:+\:\EP\cdot h_{+\mu\nu}\:+\:\EM\cdot h_{-\mu\nu}
\end{equation}
where $g_{\mu\nu}$ is the classical background metric. In this way, positive and negative-energy modes would couple only to the positive- and negative-energy modes of the metric fluctuations $h_{{\pm}\mu\nu}$. Notwithstanding questions about the consistency of this construction at the loop level, this speculation is intriguing, but further considerations along these lines are far beyond the scope of this article, and we leave them for future work.


\section{Conclusions}
\label{sec:conclusions}
 
We have described scalar quantum field theories constructed over the ring of bicomplex numbers, which incorporate both positive- and negative-energy states. We achieve reduced sensitivity to vacuum fluctuations, while, at the same time, preserving the probabilistic interpretation and eliminating the possibility of negative-energy cascades. It is tempting to speculate that the theories permitted by this construction offer intriguing possibilities for the cosmological constant and hierarchy problems, but it remains to be seen whether these theories are consistent with perturbative unitarity. This construction is readily generalized to vector bosons and spinor fields, and this will be presented elsewhere.


\section*{Acknowledgements}

This work is based, in part, on the masters dissertation of MEL~\cite{MEL} submitted to the University of Nottingham (2017) under the supervision of PM. The work of PM is supported by STFC Grant No.~ST/L000393/1 and a Leverhulme Trust Research Leadership Award. PM would like to thank Robert Dickinson and Jeff Forshaw for their earlier collaboration and their reading of the manuscript, as well as Clare Burrage, Ed Copeland, Hans-Thomas Elze, Cohl Furey and Antonio Padilla for helpful discussions. 


\end{document}